\documentclass[%
 reprint,
 amsmath,amssymb,
 aps,
]{revtex4-1}

\usepackage{graphicx}
\usepackage{dcolumn}
\usepackage{bm}

\usepackage{datetime}
\usepackage{amsmath}
\usepackage{amsthm}
\usepackage{todonotes}

\usepackage{tikz}
\usetikzlibrary{arrows, arrows.meta}

\tikzset{
BasicNode/.style={circle, draw= black!50, fill=colora!40, thin, minimum size
  = 5mm, inner sep = 0mm}
  }

\definecolor{colora}{RGB}{0,115,179}
\definecolor{colorb}{RGB}{230,154,0} 
\definecolor{colorc}{RGB}{0,154,128} 
\definecolor{colord}{RGB}{205,10,179}
\definecolor{colore}{RGB}{255,32,0}
\definecolor{colorf}{RGB}{240,228,66}
\definecolor{colorg}{RGB}{90,179,230}
\definecolor{colorh}{RGB}{205,154,179}

\newcommand\diffm[3]{\frac{\text{d}^{#1} #2}{\text{d}#3^{#1}}}

\newcommand\ave[1]{\left \langle #1 \right \rangle}
\newcommand\F{\mathcal{F}}
\renewcommand\S{\mathcal{S}}
\newcommand\U{\mathcal{U}}

\newcommand\Li{\operatorname{Li}}
\newtheorem{property}{Property}
\newtheorem{theorem}{Theorem}
\newtheorem{corollary}{Corollary}

\begin{document}

\preprint{APS/123-QED}

\title{An elementary proof of the total progeny size of a birth-death process, with application to network component sizes}

\author{Joel C. Miller}
 \email{joel.c.miller.research@gmail.com}
\affiliation{%
 Institute for Disease Modeling\\
 Bellevue, WA, USA
}%

\date{\today \currenttime}

\begin{abstract}
We revisit the size distribution of finite components in infinite Configuration Model networks.  We provide an elementary combinatorial proof about the sizes of birth-death trees which is more intuitive than previous proofs.  We use this to rederive the component size distribution for Configuration Model networks.  Our derivation provides a more intuitive interpretation of the formula as contrasted with the previous derivation based on contour integrations.  We demonstrate that the formula performs well, even on networks with heavy tails which violate assumptions of the derivation.  We explain why the result should remain robust for these networks.

\end{abstract}

\pacs{Valid PACS appear here}
\maketitle

We consider the size distribution of birth-death processes with a specific application to the sizes of components in large random networks.  An expression for the size-distribution of small connected components in Configuration Model networks was derived in~\cite{newman2007component}.  The probability a randomly chosen node is part of a component of size $n$ is
\begin{equation}
\label{eqn:newman}
\pi_n = \frac{\ave{K}}{(n-1)!} \left[\diffm{n-2}{}{z} \left[\frac{\psi'(z)}{\ave{K}}\right]^n\right|_{z=0}
\end{equation}
where $\psi(x) = \sum P(k) x^k$ is the probability generating function (PGF) of the degree distribution.  
In other words, $\pi_n$ is $\ave{K}/(n-1)$ times the coefficient of $z^{n-2}$ in $[\psi'(z)/\ave{K}]^n$.

The derivation of Eqn.~\ref{eqn:newman} required a recursive expression involving PGFs, applying Cauchy's integral formula to that expression, performing some substitutions within the integral, and then applying Cauchy's integral formula in the opposite direction~\cite{newman2007component}.  It is unsatisfying to have a simple expression whose derivation is somewhat opaque.  That is, when we have a simple expression for some physical quantity, it is usually useful to interpret the parts of that expression physically, but with existing derivations, the physical interpretation is unclear.  In this paper, we provide an alternate derivation of a well-known related theorem for the total progeny of a birth-death process, and then adapt this proof to the component size distribution.

\section{Preliminary definitions and properties}
\label{sec:properties}
We consider a birth-death process in which each individual produces
some (non-negative integer) number of offspring $m$ chosen from a
given distribution having probability generating function $\mu(z) =
\sum r_s z^s$ where $r_s$ is the probability of $s$
offspring. 
 
We assume that in the first step of the process there are $Y_0=k$
individuals, and in each subsequent generation there are $Y_i$
individuals.  We define $Z = Y_0 + Y_1 + \cdots$ (with $Z=\infty$ if the process never dies out).  We  refer to $Z$ as the ``progeny size'' (which includes the initial $k$ individuals).

We refer to the rooted tree formed by taking an initial individual and
adding edges to its offspring and edges from its offspring to their
offspring recursively as a ``birth-death tree''.  If our process
begins with $k$ initial individuals, then we have a ``birth-death
forest'' made up of $k$ birth-death trees.

For each individual in a birth-death tree, we order its offspring
(randomly) from left to right.  We similarly order the roots of each
tree in a forest.  The resulting forest of trees with the given order is a ``planted planar forest'', and the order of a depth-first traversal is
uniquely determined.  If any tree is infinite, our
sequence is infinite and some nodes may never be reached in the
traversal.  This will not affect our
proofs.\footnote{We are interested in properties of finite forests.
  If the forest is infinite, that is the only thing we need to know
  about it, so the unlabeled nodes are not important to us.}

\begin{figure}
\begin{center}
\begin{tikzpicture}
\node[BasicNode, minimum size = 4mm, fill = colorb!50] (node0) at (2.5,0) {$A$};
\node[BasicNode, minimum size = 4mm, fill = colorb!50] (node1) at (0.5,-0.8) {$B$};
\node[BasicNode, minimum size = 4mm, fill = colorb!50] (node2) at (2.5,-0.8) {$C$};
\node[BasicNode, minimum size = 4mm, fill = colorb!50] (node3) at (2.0,-1.6) {$D$};
\node[BasicNode, minimum size = 4mm, fill = colorb!50] (node4) at (1.75,-2.4000000000000004) {$E$};
\node[BasicNode, minimum size = 4mm, fill = colorb!50] (node5) at (2.25,-2.4000000000000004) {$F$};
\node[BasicNode, minimum size = 4mm, fill = colorb!50] (node6) at (3.0,-1.6) {$G$};
\node[BasicNode, minimum size = 4mm, fill = colorb!50] (node7) at (4.5,-0.8) {$H$};
\node[BasicNode, minimum size = 4mm, fill = colorb!50] (node8) at (4.5,-1.6) {$I$};
\node[BasicNode, minimum size = 4mm, fill = colorb!50] (node9) at (6,0) {$J$};
\node[BasicNode, minimum size = 4mm, fill = colorb!50] (node10) at (6.0,-0.8) {$K$};
\node[BasicNode, minimum size = 4mm, fill = colorb!50] (node11) at (5.5,-1.6) {$L$};
\node[BasicNode, minimum size = 4mm, fill = colorb!50] (node12) at (6.5,-1.6) {$M$};
\draw [-{Latex[length=2mm,width=2mm,angle'=30]}] (node0) -- (node1);
\draw [-{Latex[length=2mm,width=2mm,angle'=30]}] (node7) -- (node8);
\draw [-{Latex[length=2mm,width=2mm,angle'=30]}] (node2) -- (node6);
\draw [-{Latex[length=2mm,width=2mm,angle'=30]}] (node10) -- (node11);
\draw [-{Latex[length=2mm,width=2mm,angle'=30]}] (node10) -- (node12);
\draw [-{Latex[length=2mm,width=2mm,angle'=30]}] (node0) -- (node7);
\draw [-{Latex[length=2mm,width=2mm,angle'=30]}] (node9) -- (node10);
\draw [-{Latex[length=2mm,width=2mm,angle'=30]}] (node2) -- (node3);
\draw [-{Latex[length=2mm,width=2mm,angle'=30]}] (node3) -- (node4);
\draw [-{Latex[length=2mm,width=2mm,angle'=30]}] (node0) -- (node2);
\draw [-{Latex[length=2mm,width=2mm,angle'=30]}] (node3) -- (node5);
\end{tikzpicture}\\[20pt]
$\mathcal{U} = (A, B, C, D, E, F, G, H, I, J, K, L, M)$\\
$\mathcal{S} = (3, 0, 2, 2, 0, 0, 0, 1, 0, 1, 2, 0, 0)$
\end{center}
\caption{A forest $\F{}$ and the corresponding sequences $\U{}$ and
  $\S{}$ from the depth-first traversal in $\phi(\F)$.}
\label{fig:forest_to_sequence}
\end{figure}

We consider a planted planar forest $\F{}$ with $k$
trees.  We define the mapping $\phi$ so that $\phi(\F{})$
produces a sequence $\U_{\F}=(u_0, u_1, \ldots, u_{Z-1})$ of the nodes in the
depth-first traversal order and $\S_{\F}=(s_0, s_1, s_2, \ldots, s_{Z-1})$
where $s_i$ is the number of offspring of $u_i$ (allowing that the
sequences may be infinite).  An example of $\phi$ is in
Fig.~\ref{fig:forest_to_sequence}.

\begin{figure*}
\input{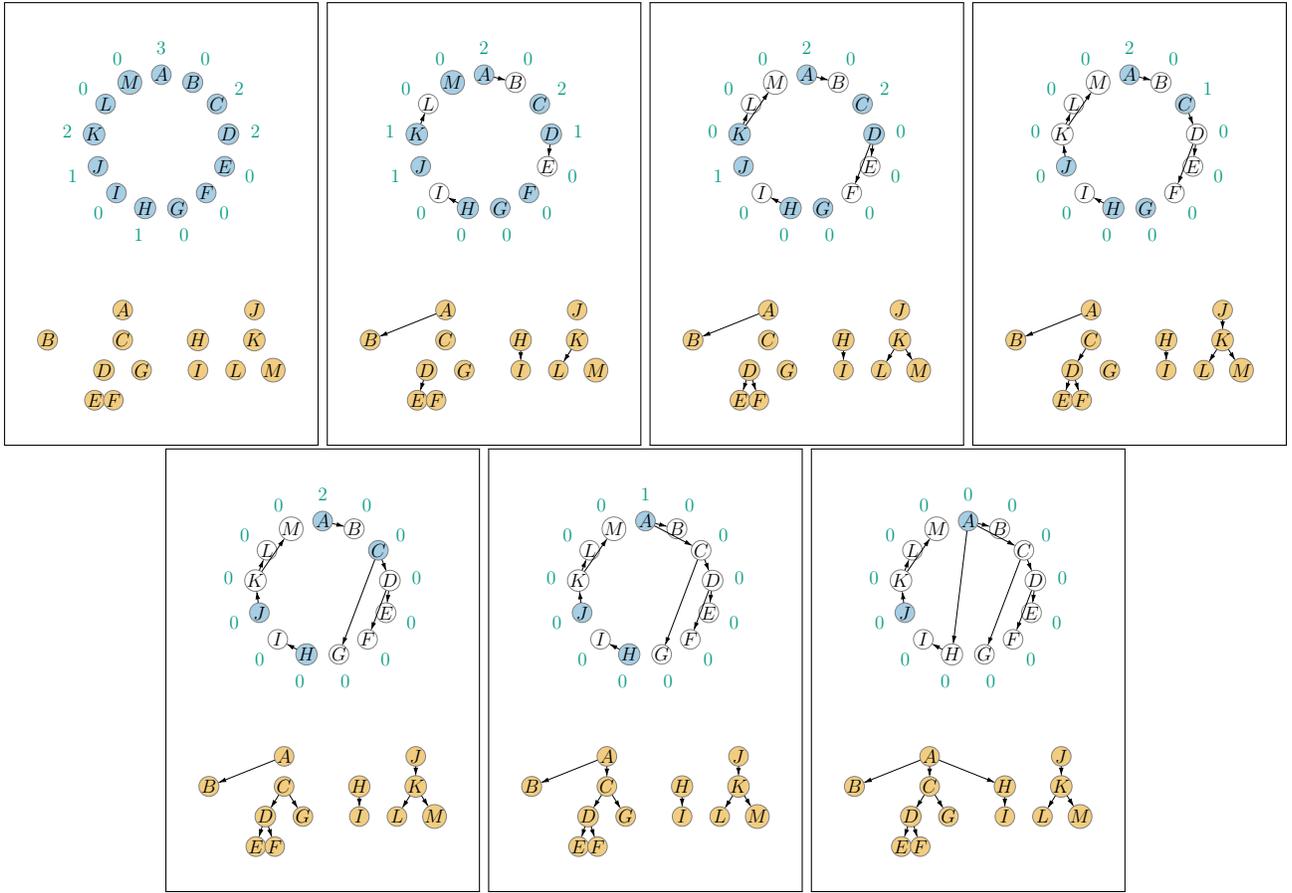}
\caption{Example of $\xi$.  $\U{}=(A, B,\ldots, M)$ and
  $\S{} = (3, 0, 2, 2, 0, 0, 0, 1, 0, 1, 2, 0, 0)$ (corresponding to Fig.~\ref{fig:forest_to_sequence}).  In the ring, the
  black number inside the node is the node label, while the colored
  number beside the node is the value of $s_i$ at that step.  Nodes in
  the ring are filled if they have not yet been ``processed''.   This yields a
  forest $\F{}$ with two trees, one rooted at $u_0=A$.  Here
  $(\U{},\S{})=\phi(\F{})$.  In showing how the tree is wired
  together, we pre-place the nodes in the correct position.}
\label{fig:sample_algorithm1}
\end{figure*}

\begin{figure*}
\input{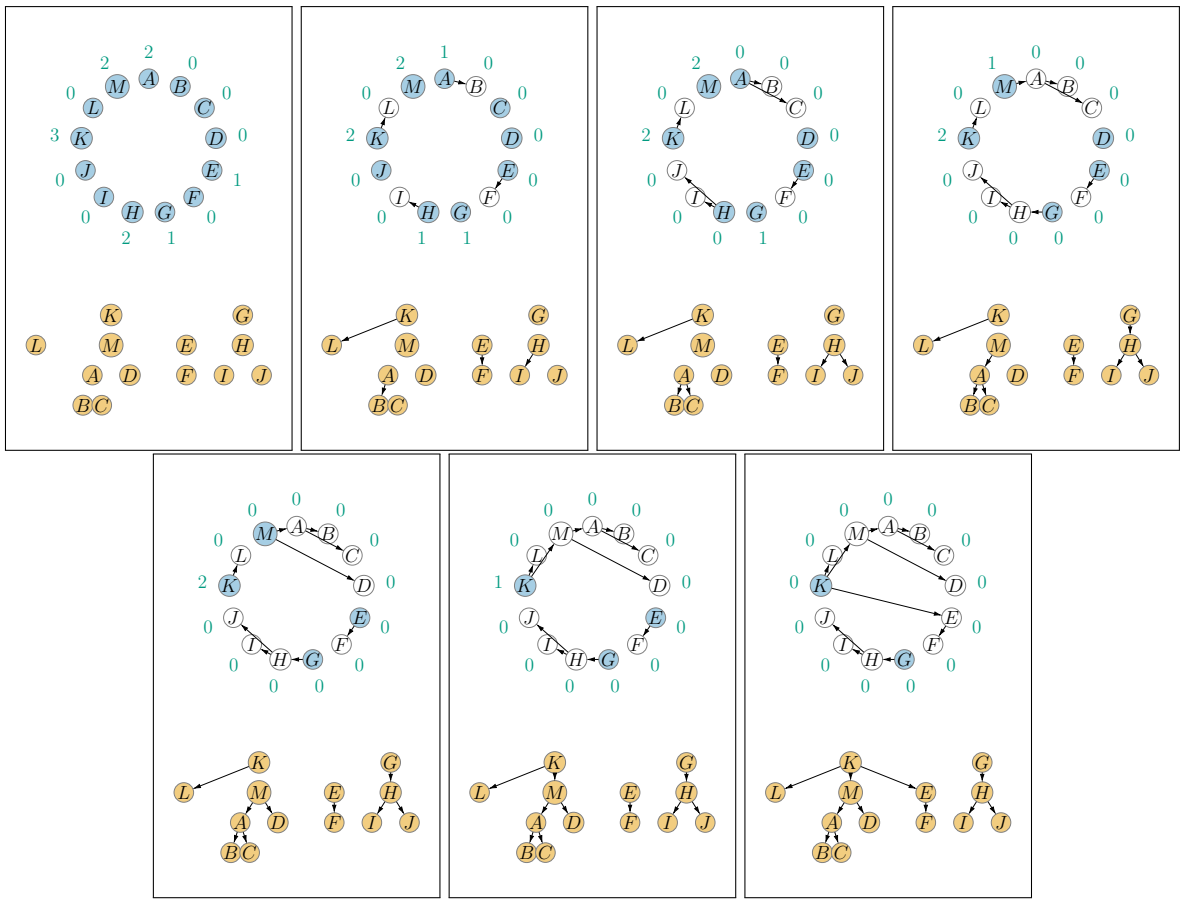}
\caption{Another example of $\xi$.  $\U{}=(A, B,\ldots,M)$ 
  as in Fig.~\ref{fig:sample_algorithm1}, but $\hat{\S}$ is a cyclic permutation of $\S{}$ to
  $(2, 0, 0, 0, 1, 0, 1, 2, 0, 0, 3, 0, 2)$.  This yields a forest
  $\hat{\F}$ with two trees, but neither tree is rooted at $u_0=A$.
  The shape of $\hat{\F}$ is the same as $\F{}$, but the labels have
  undergone a cyclic permutation.}
\label{fig:sample_algorithm2}
\end{figure*}

We will prove properties of a birth-death
forest $\F{}$ by investigating properties of the sequence
$\S_{\F}$.  First we note that if $\F{}$
is finite  then $\sum s_i = Z-k$ because $Z = k + \sum s_i$.

Given a finite sequence of $n$ non-negative integers
$\S{} = (s_1, \ldots, s_{n-1})$ with sum $n-k$ and a finite sequence
of nodes $\U{} = (u_1, \ldots, u_{n-1})$ we define a mapping below,
$\xi(\U{},\S{})$ that creates a forest of $k$ trees.  In those cases
where there is a forest for which $(\U{},\S{}) = \phi(\F{})$, we will
see that $\xi=\phi^{-1}$.  That is $\F{}=\xi(\phi(\F))$ for all finite
birth-death forests $\F{}$.

However, there are examples of $\S{}$ which cannot result from a
birth-death forest.  For example, given a depth-first traversal of a
forest, we are guaranteed that the final node visited has $0$
offspring.  Thus if $s_{n-1}\neq 0$ then $\mathcal{S}$ does not
correspond to a birth-death forest.

We define $\xi$ algorithmically and demonstrate it in
Figs.~\ref{fig:sample_algorithm1} and~\ref{fig:sample_algorithm2}.
Given an arbitrary finite sequence $\S{}$ of $n$ non-negative integers
whose sum is $n-k$ and an ordered sequence of nodes:
\begin{itemize}
\item We place the nodes $u_0, \ldots, u_{n-1}$ into a ring and
  mark all nodes as ``unprocessed''.
\item While it is possible to find at least one unprocessed node $j$
  with $s_j=0$ such that the previous unprocessed node in the ring $i$
  has $s_i \neq 0$, we repeat the following steps:
\begin{enumerate}
\item Find all unprocessed nodes $u_{j_1}, \ldots, u_{j_L}$ for which
  $s_{j_\ell}=0$ and for which the previous unprocessed nodes in the ring $u_{i_1},
  \ldots, u_{i_L}$ have $s_{i_\ell}>0$ (note that there is no node that appears both as one of the $u_j$ and one of the $u_i$ nodes).
  \item We put $u_{j_\ell}$ into the left-most available offspring position for
  $u_{i_\ell}$.  \label{adding_step}
\item We mark each $u_{j_\ell}$ as processed and remove them from the ring
\item We reduce each $s_{i_\ell}$ by one.
\item We repeat with the remaining ring. \label{recursive_step}
\end{enumerate}
\item The resulting forest is defined to be $\xi(\U{},\S{})$.
\end{itemize}
This process has several properties, which we prove in the supplement.

Note that we could define $\xi$ recursively by simply moving step~\ref{adding_step} to be after step~\ref{recursive_step}.  The ring of unprocessed nodes that go through the next iteration would be the same in both formulations (resulting in the same edges added in subsequent steps), and in both cases $j_\ell$ would become the left-most offspring of $i_\ell$.  

\begin{property}
$\xi(\U,\S)$ is a birth-death forest with $k$ trees.
\end{property}

\begin{property}
If $u_i$ is a root of $\F=\xi(\U,\S)$ and we perform a cyclic
permutation to create $\hat{\U} = \sigma_i(\U)$ and $\hat{\S} = \sigma_i(\S)$ so that 
\begin{align*}
\hat{\U} &=
(u_i, u_{i+1}, \ldots, u_{n-1}, u_0, u_1, \ldots, u_{i-1})\\
\hat{\S} &=
(s_i, \ldots, s_{n-1}, s_0, \ldots, s_{i-1})
\end{align*}
then if the roots of $\F$ are ordered as they are in $\hat{\U}$ we
have
\[
(\hat{U},\hat{S}) = \phi(\F)
\]
\end{property}

These properties establish that if $u_0$ is a root of
$\F=\xi(\U,\S)$ then a depth-first search of $\F$ that records the
nodes and their number of offspring will produce $\U$ and $\S$.  
Further, given a sequence $\S$ summing to $n-k$ then exactly $k$ of
the $n$ permutations make the first element of $\hat{\U}$ into a root.
For these cyclic permutations (and no others) there is a forest $\F$
such that $\hat{U}$ and $\hat{S}$ correspond to a depth-first search
of $\F$.




\section{The total progeny size}

We prove the following result~\cite{dwass1969total}:
\begin{theorem}
\label{thm:birthdeath}
Given a birth-death process starting with $k$ individuals where each
individual produces a non-negative number of offspring from some
imposed distribution, the probability the total progeny size $Z$ satisfies $Z=n$ is $k/n$ times the probability that $n$ numbers chosen
from that distribution would sum to $n-k$.
\end{theorem}

In general, previous proofs of this result rely on PGFs and contour integration.  Our proof
will simply use the properties of $\phi$ and $\xi$ described above.
The gist of the proof is that there is a one-to-one correspondence
from forests $\F$ to sequences $\U$ and $\S$.  We will use the fact
that for a given random sequence exactly $k$ of the $n$ cyclic
permutations correspond to trees to show that the probability a random
sequence corresponds to a forest is $k/n$.  With a few additional
technical details, we then show that the probability a
forest has $n$ nodes is $k/n$ times the probability a random length-$n$
sequence sums to $n-k$.

\begin{proof}
We assume that both $n$ and $k$ are given.


We consider a planted planar forest $\F{}$ started from $k$
individuals, and we define $\U$ and $\S$ to be $(\U,\S) = \phi(\F)$.  Without loss of generality, we assume that the nodes of
$\F$ are labeled in order so that $\U=(0,1,\ldots,Z-1)$ where $Z$ is
the (possibly infinite) number of nodes in $\F$.  The probability that
$\F$ has a particular shape is
$\pi_\F = \prod_{s_i \in \S_{\F}} r_{s_i}$ where $r_s$ is the
probability of having $s$ offspring.

Our goal is to find the probability $\pi_n$ of having size $n$, where
$n$ is finite.  This
is
\[
\pi_n = \sum_{\F: |\F|=n} \pi_\F
\]

On the other hand, if we choose a sequence of $n$ numbers $\S$ where each
number is chosen from the offspring distribution, the probability of
choosing a particular $\S$ is $\pi_\S = \prod_{s_i \in \S} r_{s_i}$.
So $\pi_\F = \pi_{\S_\F}$.

We can now focus on the easier probability space consisting of
sequences of length $n$.  It is clear that $\pi_n = \sum \pi_\F$ is
equal to the sum of $\pi_\S$ taken over those sequences $\S$ for which
there exists an $\F$ with $\S_\F=\S$.  Our goal now is to find the
probability a given a random
sequence $\S$ of length $n$ there exists an $\F$ with $(\U,\S) = \phi(\F)$.

We consider now a randomly chosen sequence $\S$ that sums to $n-k$
elements.  We collect all the cyclic permutations of $\S$ and put them
into an equivalence class $C$.  All of these sequences have the same
probability.  By the properties in Section~\ref{sec:properties}, a fraction $k/n$
of these cyclic permutations correspond to a forest $\F$.  We define
$\pi_C = |C|\pi_S$ to be the probability a random length-$n$ sequence is in $C$.
So the probability a random sequence is in $C$ and corresponds to a planted planar forest is $\pi_C k/n$.

We now partition all  length $n$ sequences which sum to
$n-k$ into a finite number of disjoint equivalence
classes $C_1, C_2, \ldots C_A$.  Two sequences are in the same
equivalence class $C_\alpha$ if and only they are cyclic permutations
of one another.  
\begin{align*}
\pi_n &= \sum_{\F: |\F|=n} \pi_{\F}\\
 &= \sum_{\S_{\F}: |\F|=n} \pi_{\S_{\F}}\\
&= \sum_\alpha \frac{k}{n} \sum_{C_\alpha} \pi_{C_\alpha}\\
&= \frac{k}{n} \sum_{\S: s_0+\cdots +s_{n-1}=n-k} \pi_{\S}\\
\end{align*}
The final equality results from the fact that every sequence which
sums to $n-k$ is in exactly one $C_\alpha$, so $\sum \pi_{C_\alpha} =
\sum \pi_{\S}$ where the first summation is over all equivalence classes
that sum to $n-k$ and the second summation is over all sequences that
sum to $n-k$.
\end{proof}

As a technical point, we note that if $k$ and $n$ are not relatively
prime, different equivalence classes may have a different number of
sequences.  For example: both $C_1=\{(1,0,1,0), (0,1,0,1)\}$ and
$C_2 = \{(2,0,0,0), (0,2,0,0), (0,0,2,0), (0,0,0,2)\}$ are equivalence
classes with $n=4$ and $k=2$.  The probability $0$ is a root in the
resulting forest is still $k/n$ for both.

The theorem can be interpreted in the following way: given a sequence
of $n$ non-negative integers that sum to $n-k$ arranged on a ring the
sequence encodes the degrees found in a depth-first traversal of $k$ trees.  Specific positions in the sequence correspond to the roots of those $k$ trees.
As we rotate that sequence around the ring there are $n$ possible
rotations, exactly $k$ of which result in a root at the top.  These
are the only sequences we want, and thus the probability a random
sequence that sums to $n-k$ comes from an planted planar forest is
$k/n$.    Thus the probability that the length-$n$
sequence forms a planted planar forest equals $k/n$ times the probability
that the sequence sum to $n-k$.

We can re-express Theorem~\ref{thm:birthdeath} in terms of probability generating functions.  
\begin{corollary}
\label{cor:birthdeath}
Consider a birth-death process beginning with $k$ individuals.  If $\mu(z)$ is the probability generating function of the offspring distribution, then the probability of exactly $n$ progeny is the coefficient of $z^{n-k}$ of $\frac{k}{n}[\mu(z)]^n$.
\end{corollary}
This is proven by noting that the coefficient of $z^i$ in $[\mu(z)]^n$ is the probability that $n$ numbers chosen from the distribution sum to $i$.  





\section{Component size distribution of Configuration Model networks}
We now look at the component size distribution  of a large
Configuration Model network.  We initially assume that the degree distribution
has finite second moment, so that for a given $n$, the probability an
a randomly chosen node is in a short cycle scales like $1/N$ as $N \to \infty$.  That is, we assume the network is locally tree-like.  The probability of choosing a node with degree $k$ is $p_k$.  We define the PGF $\psi(z) = \sum p_kz^k$.  If we consider the random neighbor of a node, the probability the neighbor has degree $\hat{k}$ is $\hat{k} p_{\hat{k}}/\ave{K}$ where $\ave{K}$ is the average degree.  The so-called \emph{excess degree} of the neighbor is the number of edges other than the edge it was reached along, $\hat{k}-1$.  The PGF of the excess degree distribution is $\sum_{\hat{k}} \hat{k}p_{\hat{k}} z^{\hat{k}-1}/\ave{K} = \psi'(z)/\ave{K}$.

We seek to calculate the size distribution of the component containing
a randomly chosen node $u$.  We take $u$ to have degree $k$.  We
remove $u$ from the network and look for the the sum of the sizes of
the components containing its $k$ neighbors.  We seek the probability
that the component including $u$ has size $n$, so we look for the
probability the components with $u$ removed have size $n-1$.  

Each tree started from a neighbor of $u$ corresponds to a birth-death process with offspring distribution chosen from the excess degree distribution.  From Corollary~\ref{cor:birthdeath} the  probability that they sum to $n-1$ is given by the coefficient of $z^{n-1-k}$ in $\frac{k}{n-1}[\psi'(z)/\ave{K}]^{n-1}$.   Thus if $u$ has degree $k$, the probability the component including $u$ has size $n$ is the coefficient of $z^{n-2}$ in $z^{k-1}\frac{k}{s-1}[\psi'(z)/\ave{K}]^{n-1}$ (note we multiplied by $z^{k-1}$ to change the exponent of the term whose coefficient we want). 

Summing over all degrees $u$ might have, we see that the probability the component including $u$ has size $n$ is the coefficient of $z^{n-2}$ of
\begin{align*}
 \sum_k P(k) z^{k-1} \frac{k}{n-1} \left[\frac{\psi'(z)}{\ave{K}}\right]^{n-1}
 &= \frac{1}{n-1} \psi'(z) \left[\frac{\psi'(z)}{\ave{K}}\right]^{n-1}\\
&= \frac{\ave{K}}{n-1} \left [ \frac{\psi'(z)}{\ave{K}}\right]^n
\end{align*}
This is identical to Eq.~\eqref{eqn:newman} because a way to choose the coefficient of $z^{n-2}$ is to take $n-2$ derivatives, divide by $(n-2)!$ and then evaluate at $z=0$.  In practice however, for many distributions it will be easier to determine the expansion and identify the correct coefficient, rather than performing the derivatives.

\subsection{Examples}

\paragraph{Poisson degree distribution}

We consider a Poisson degree distribution with mean $\lambda$.  The
PGF is $\psi(z) = e^{-\lambda (1-z)}$.  From this
\[
\psi'(z) = \lambda e^{-\lambda(1-z)}
\]
and 
\[
\psi'(z)/\ave{K} = \psi(z)
\]
The coefficient of $z^{n-2}$ in $\frac{\ave{K}}{n-1}\psi(z)^n$ is straightforward to find
using $\psi(z)^n = e^{-\lambda n(1-z)} = e^{-\lambda n}e^{\lambda n
  z}$.  By expanding $e^{\lambda n z}$ as a Taylor series we have that the probability of a
component of size $n$ is
\[
\frac{\lambda}{n-1} e^{-\lambda n} \frac{(\lambda n)^{n-2}}{(n-2)!} =
\frac{(n\lambda)^{n-1}}{n!}e^{-\lambda n}
\]

\begin{figure}
\includegraphics[width=\columnwidth]{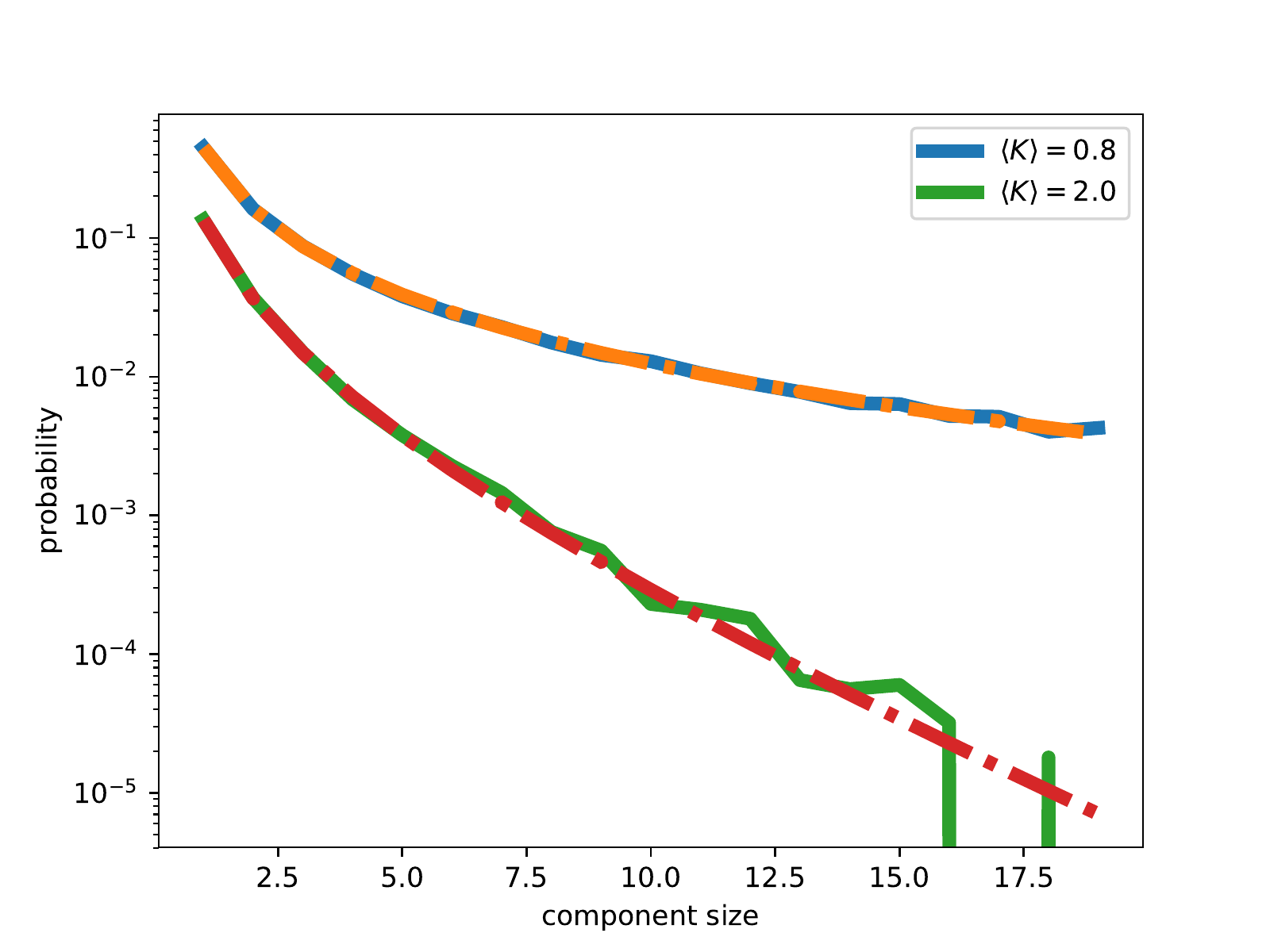}
\caption{A comparison of the predicted (dot-dashed) and observed (solid) component size frequency in a single network of $10^6$ nodes with Poisson degree distribution.}
\end{figure}

\paragraph{Power-law degree distribution}

We consider $P(k) = ck^{-3}$ for $k=1, 2, 3, \ldots$.  For this model, the expected degree $\sum c k^{-2} = c \pi^2/6$ is
finite, while the second moment $\sum c/k$ is infinite.

We can find $\psi'(z) = c\sum  k^{-2} z^{k-1}$ where $c=\left(\sum_{k=1}^\infty 1/k^3\right)^{-1} \approx 1/1.202$ is the inverse of Ap{\'e}ry's constant~\cite{apery1979irrationalite}.  Multiplying by $z$ gives $z\psi'(z) = c \sum k^{-2} z^k = c\Li_2(z)$ where $\Li_2(z)$ is a polylogarithm.  Note that $\Li_2(1) = \sum 1/k^2 = \pi^2/6$. So
\[
\psi'(z) = c \frac{\Li_2(z)}{z}
\]
and
\[
\psi'(z)/\psi'(1) = \frac{6}{\pi^2} \frac{\Li_2(z)}{z}
\]
and $c$ drops out of this expression.  

The probability of a size-$n$ component is the coefficient of $z^{n-2}$ in
\[
\frac{c\pi^2}{(n-1)6}\left(\frac{6}{\pi^2}\frac{\Li_2(z)}{z}\right)^n
=\frac{c}{n-1} \left(\frac{6}{\pi^2}\right)^{n-1}\left(\frac{\Li_2(z)}{z}\right)^n
\]
It is
straightforward to use symbolic calculation to find the first
coefficients of $[\psi'(z)/\ave{K}]^n$ for the first few values of $n$.  It becomes more difficult for larger $n$.  
We have
\[
\pi_1=0, \qquad \pi_2 = c\frac{6}{\pi^2}, \qquad \pi_3 = c\frac{27}{2\pi^4}, \qquad \pi_4 = c\frac{59}{\pi^{6}}
\]
We turn to the Cauchy integral formula for general $n$.  Given an analytic function $f(x) = \sum a_k x^i$, the coefficient can be calculated by
\[
a_k = \frac{1}{2\pi i} \oint \frac{f(z)}{z^{k+1}} \, \mathrm{d}z \, .
\]
This integral can be well-approximated by 
\[
a_k \approx \frac{1}{M}\sum_{m=0}^{M} \frac{f(Re^{ 2 \pi im/M})}{R^k e^{2k\pi i m/M}}
\]
where $R$ represents the radius of a circle in which the function is analytic.

\begin{figure}
\includegraphics[width=\columnwidth]{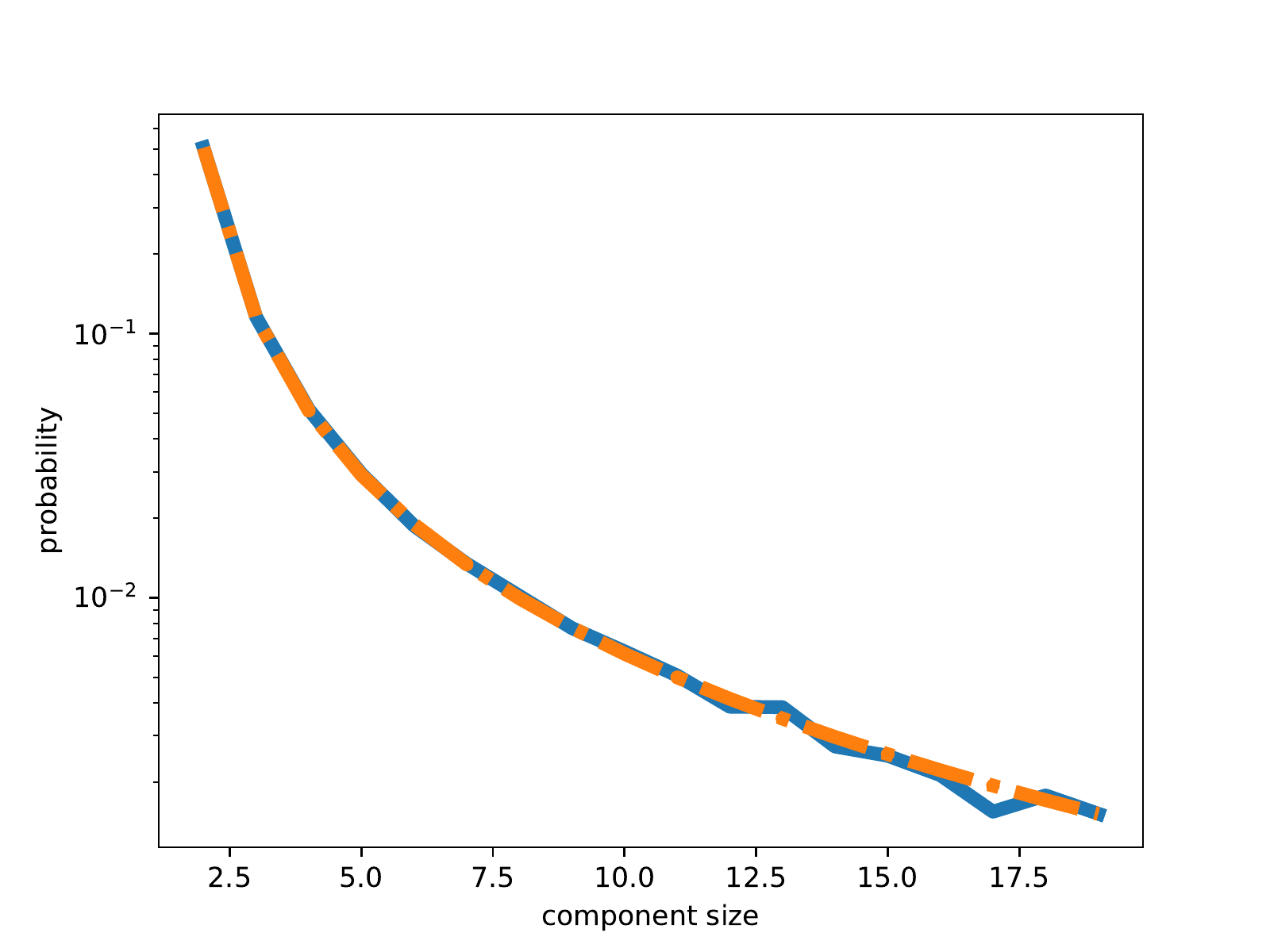}
\caption{A comparison of prediction (dot-dashed) and observation (solid) in a network of $10^6$ nodes having $P(k) = c/k^3$ for $k>0$.}
\label{fig:scalefree}
\end{figure}

\subsection{Validity of model at high variance}
Our proof was derived assuming $\ave{K^2}$ is finite, for which the
probability that a random node is part of a short cycle goes to $0$
like $1/N$.  Our first  example shows that the formula behaves well in such networks.  Our second example shows that it performs surprisingly well in networks with high variance which do not look locally tree-like in general.  

The reason for this is that the proof relies on the assumption that the small component has no cycles.  The reason that networks do not look locally tree-like when their degree distribution has high variance is that given an edge $u$--$v$, there will be high degree nodes which are likely to form a triangle with $u$ and $v$ simply by virtue of having very high degree.    

Revisiting the proof, the steps of following a birth-death process are valid until a node is put into the tree more than once.  So our question is: ``does the answer for the probability of a small component size change if we treat multiple additions of the same high-degree node as being separate additions of different high-degree nodes?''  We argue that the answer is no, because we anticipate that the first addition of the high-degree node is likely to guarantee an infinite component.

So we expect that the formula in Eqn~\eqref{eqn:newman} performs well even if the network itself is not locally treelike because it is locally treelike within the small components.

\section{Discussion}
This paper gave a new derivation of the component size distribution for Configuration Model networks, under the locally-treelike assumption.  The derivation gives a combinatorial explanation without relying on properties of contour integration.  We believe that this yields an intuitive physical explanation of the previously derived result.

Additionally, we explained why the resulting formula should apply in large networks even if the degree distribution forces a non-negligible number of short cycles.  Those short-cycles only appear in the components that are not small.
\appendix

\section{Supplement}
In this supplement we prove the properties mentioned in Section~\ref{sec:properties}, which are effectively lemmas for theorem~\ref{thm:birthdeath}.

We first show that $\xi(\U,\S)$ is a birth-death forest with $k$ trees.
\begin{proof}
At each step, when we add an edge, the edge is between two unprocessed nodes.  Upon the edge being added, one of the nodes is labeled as processed.

Arguing inductively, if there is no path between any two unprocessed nodes before an edge is added, the addition of an edge between two unprocessed nodes $u$ and $v$ cannot create a new cycle because there was no $u$--$v$ path initially, and it also does not create any new path between two unprocessed nodes other than $u$ and $v$.  By moving one of $u$ and $v$ from unprocessed to processed, we guarantee that at the next step there is still no $u$--$v$ edge.

Because of this, the result is a forest.
It has $k$ trees because at the $r$th iteration, if $n_r$ is the number of unprocessed nodes, the sum of the $s$ for those nodes is $n_r-k<n_r$.  If this is positive, then there must be an $s_i>0$ and an $s_j=0$.  Thus there is at least one pair that will have an edge added.  If the sum is zero, then there are $k$ remaining unprocessed nodes and the process stops.  Thus we have $k$ distinct connected components, which are rooted at those final $k$ nodes.
\end{proof}

We now show that if $u_i$ is a root of $\F=\xi(\U,\S)$ then by a cyclic permutation that moves $u_i$ to the first position, we get a pair $\hat{U}$ and $\hat{S}$ such that $(\hat{U},\hat{S}) = \phi(\F)$.  

\begin{proof}
If the process results in the top node of the ring being a root, then it is clear that $\U$ and $\S$ satisfy $(\U,\S) = \xi(\F)$ where $\F = \phi(\U,\S)$ with the ordering of the trees being as they appear in $\U$. 

If there is a root which is not at the top of the ring, then rotating the ring so that the root does appear at the top corresponds to performing a cyclic permutation of $\U$ and $\S$.  The resulting tree remains the same because the steps adding edges only care about relative position in the ring.  Once this is done, we are back in the situation where the root is at the top.
\end{proof}

\bibliography{component_size}
\end{document}